\documentclass[11pt,letterpaper,aps,preprint,groupedaddress,showpacs,amsmath,amssymb]{revtex4-1}
\usepackage[margin=2.25cm]{geometry}
\usepackage{setspace}
\DeclareRobustCommand{\rchi}{{\mathpalette\irchi\relax}}
\newcommand{\irchi}[2]{\raisebox{\depth}{$#1\chi$}}
\usepackage{graphicx}
\usepackage{dcolumn}
\usepackage{bm}
\begin{document}

\title{The fundamentals of harnessing the magneto-optics of quantum wires for designing optical
       amplifiers: Formalism}
\author{Manvir S. Kushwaha\\
Department of Physics and Astronomy, Rice University, P.O. Box 1892, Houston, TX 77251, USA}

\bigskip

\date{\today}
\begin{abstract}
Quantum wires occupy a unique status among the semiconducting nanostructures with reduced dimensionality
-- no other system seems to have engaged researchers with as many appealing features to pursue.
This paper aims at a core issue related with the magnetoplasmon excitations in the quantum wires
characterized by the confining harmonic potential and subjected to a longitudinal electric field and a
perpendicular magnetic field in the symmetric gauge. Despite the substantive complexity, we obtain the
exact analytical expressions for the eigenfunction and eigenenergy, using the scheme of ladder operators,
which fundamentally characterize the quantal system. Crucial to this inquiry is an
intersubband collective excitation that evolves into a magnetoroton -- above a threshold value of magnetic
field -- which observes a negative group velocity between the maxon and the roton. The evidence of negative
group velocity implies anomalous dispersion in a gain medium with the population inversion that forms the
basis for the lasing action of lasers. Thus, the technological pathway that unfolds is the route to devices
exploiting the magnetoroton features for designing the novel optical amplifiers at nanoscale and hence
paving the way to a new generation of lasers.
\end{abstract}

\maketitle

\newpage

\section{Introduction}

Communication is broadly defined as the transfer of information from one point to another using
mutually understood semiotic rules. In optical fiber communications, this transfer is achieved
by using light as the global information carrier. Opto-electronic technology that has been at
the forefront of the early development of optical communications is also central to the
realization of the future networks that will have the capabilities to meet the growing demands
of mankind. These capabilities include practically unlimited bandwidth to transport communication
of almost any kind. Our target bearer is the system of quantum wires.

The quantum wires lie in the middle of the quantum rainbow comprising the quasi-$N$ dimensional
electron systems (Q-$N$DES) of diminishing dimensions -- with $N$ ($\le 2$) being the degree of
freedom. The proposal of the quantum wire structures was motivated by the suggestion [1] that 1D
k-space restriction would severely reduce the impurity scattering, thereby substantially
enhancing the low-temperature electron mobilities. As a result, the technological promise that
emerges is the route to the faster opto-electronic devices fabricated out of quantum wires. The
quantum wires have exhibited some singular properties such as the electron waveguide, magnetic
depopulation, quantization of conductance, spin-charge separation, quenching of the QHE, quantum
pinch effect [2], ...etc. These distinctions, which have been systematically discussed in Ref. 3,
motivated us to explore the quantum wires for their role as the novel gain medium in designing
semiconducting optical amplifiers.

Many strides have been made in optical networks since the advent of optical amplifiers. An optical
amplifier is a device that amplifies an optical signal directly -- without the need of optical to
electrical to optical conversions [4]. They can be divided into two categories: doped fiber
amplifiers (DFAs) and semiconducting optical amplifiers (SOAs) -- both marked by certain
{\it pros and cons} related with, e.g., the temperature and/or polarization dependence. Most
realistic applications -- in signal generation and propagation -- suffer from the innate
attenuation losses characteristic of the medium and its geometry. Thanks to the advances in
nanofabrication technology and electron lithography, the emergence of semiconducting
nanostructures with reduced dimensionality has made SOAs the most versatile technology [5].
Employing a semiconducting nanostructure with dimensional constraint (on the motion of charge
carriers) has a huge impact on the device properties. This has stimulated an upsurge in devising
nanowire-based [6], quantum-well-based [7], and quantum-dot-based [8] SOAs in the past two decades
[9]. To the best of our knowledge, no quantum-wire-based SOAs have been investigated thus far, and
this is the goal of the present work.

Essentially, we focus on the evolution of an otherwise regular intersubband collective excitation (ICE)
into a magnetoroton (MR). In the quantum wires, a magnetoroton is an ICE mimicking the roton-like
spectrum and its existence is {\it solely} attributed to the applied magnetic field ($B$) in the system.
A roton is, by definition, an elementary excitation (or a quasi-particle) whose linear rise from the
origin, the maximum, and the minimum -- in energy as the momentum increases -- are termed, respectively,
the phonons, the maxons, and the rotons (see, e.g., Fig. 1). The term roton was coined for the vortex
spectrum ($\nabla \times {\bf V}_g \ne 0$) -- ${\bf V}_g$ being the group velocity in the quantum
liquid -- of elementary excitations in the superfluid He II by Landau in 1941. The MR changes the
sign of its group velocity {\it twice} before merging with the respective single-particle continuum. As
is demonstrated in what follows, there is a minimum (threshold) value of B ($B_{th}$) and the ICE evolves
into an MR if and only if $B \ge B_{th}$. The MR in quantum wires was predicted within the framework of
Hartree-Fock approximation [10] and soon (partially) evidenced in the Raman scattering experiments [11].
A {\it rigorous} finding of the MR in the realistic quantum wires had, however, been elusive [12-13]
until 2007 [14].
\begin{figure}[htbp]
\includegraphics*[width=7.5cm,height=7.5cm]{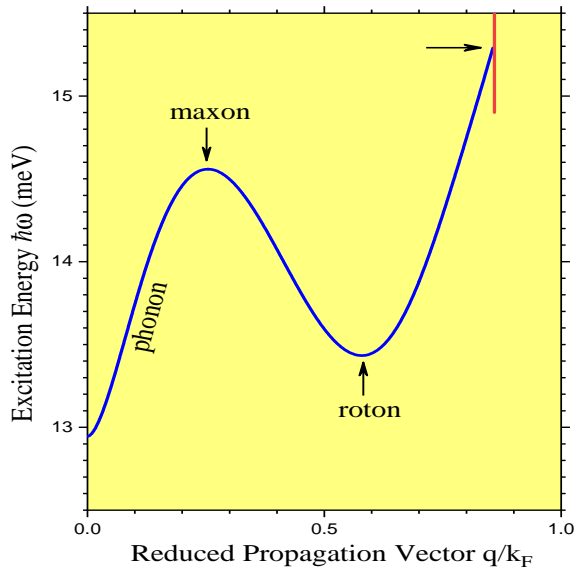}
\centering
\caption{(Color Online) The MR marked $\Omega_{01}$ in Fig. 4. The red line refers to the point where
the MR merges with the (intersubband) single-particle continuum.}
\label{fig1}
\end{figure}

The rest of the article is organized as follows. In Sec. II, we present the theoretical framework leading
to the derivation of the eigenfunction and the eigenenergy characterizing the system of quantum wires.
This is followed by the derivation of the nonlocal, dynamic, dielectric function, which is further
diagnosed analytically to fully address the solution of the problem and the associated relevant aspects.
In Sec. III, we discuss several illustrative examples of, for instance, the density of states, variation
of the Fermi energy and subband occupation, the excitation spectrum comprising of single-particle and
collective excitations, the group velocity of the MR excitations, the gain coefficient, and the life-time
of the MR excitations in such quantum systems. Finally, we summarise our finding with the specific remarks
regarding the interesting features worth adding to the problem in Sec. IV.

\section{Methodological Formalism}

The model quantum wire investigated in the present work is depicted in Fig. 2. We start with a Q-2DES in
the x-y plane subjected to a harmonic confinement potential $V(x)=\frac{1}{2}\,m^* \,\omega_o^2 \,x^2$
along x direction, a strong (quantizing) magnetic field in the symmetric gauge specified by the vector
potential ${\bf A}=\frac{1}{2}\,B$ (-$y$, $x$) [see Appendix A], and a weak (non-quantizing) electric
field $E_y$ along the -y direction. The resultant system is a quantum wire (or, more realistically, a
Q-1DES for better and broader range of physical understanding) with free propagation along the y
direction and the magneto-electric quantization along the x direction. This quantum wire is distinguished
by the single-particle [of charge $-e$, with $e>0$] Hamiltonian
\begin{figure}[htbp]
\includegraphics*[width=8.1cm,height=8.1cm]{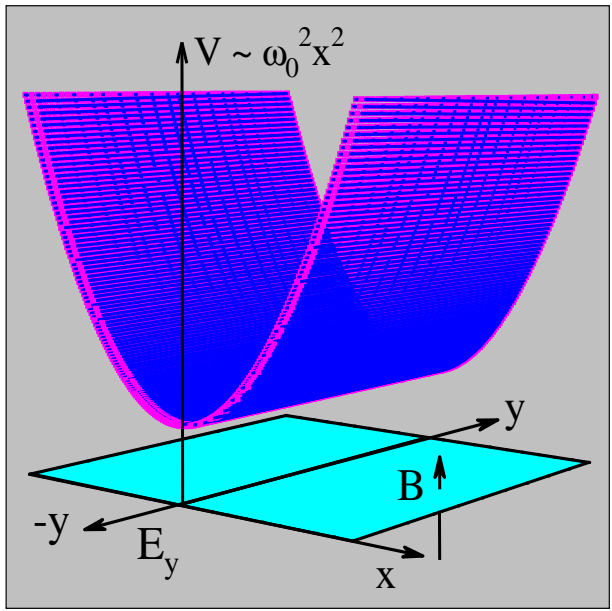}
\centering
\caption{(Color Online) The effective geometry of the model quantum wire investigated in this work.}
\label{fig2}
\end{figure}
\begin{align}
H=\frac{1}{2m^*}\,\Big[\Big(p_x + \frac{e}{c}A_x\Big)^2 + \Big(p_y + \frac{e}{c}A_y\Big)^2\Big]
                            +\, \frac{1}{2}\,m^* \,\omega_o^2 \,x^2 +e E_y y\, ,
\end{align}
in the Coulomb gauge, where $\omega_o$ is the characteristic frequency of the harmonic oscillator. Note
that $E_y$ merely serves to speed up the electrons along the degree of freedom. The rest of the symbols
have their usual meanings.

\subsection{Eigenfunction and eigenenergy}

The unceasing failures in the efforts for obtaining the exact eigenfunction and eigenenergy of the system
within the {\it wave mechanics} compelled us to pursue the {\it method of second quantization}, which did
enable us to determine the exact eigenfunction
\begin{align}
\psi_{n\alpha k}(x,y)=\frac{N_n}{\sqrt{L_y}}\,\Big\{\Big[\Big(\frac{x'}{2}+C^+\Big)-&\partial_{x'}\Big]^n\,
                        e^{-(\frac{x'}{2}+C)^2}\Big\}\, e^{(C^2-K_l^2)}\, \nonumber\\
                      &\times\, e^{-i(\alpha l_e+\frac{y''}{2})x'}\,e^{iky}\,
\end{align}
and the eigenenergy
\begin{align}
\epsilon_{n \alpha k}=\Big(n+\frac{1}{2}\Big)\,\hbar \tilde{\omega} + \frac{\hbar^2 k^2}{2 m_r} -
               \lambda_o \,\frac{\hbar \alpha}{m^*} - \frac{\lambda_o^2}{2m^*}\,
\end{align}
of the system represented by the Hamiltonian in Eq. (1). {\it The middle brackets in 
Eq. (2) limit the range of the operator $\partial_{x'}$($=\partial/\partial x'$) in the square brackets.
Ignoring this subtlety
can lead one to obtain erroneous results}. Here
$N_n={\rm [}\sqrt{\pi}\, n!\, 2^n\,(\sqrt{2}\,l_e){\rm ]}^{-1/2}$
is the normalization coefficient of the x-dependent part and $\sqrt{L_y}$ is the normalization length of
the y-dependent part of the eigenfunction. Other symbols are defined as
follows: $\tilde{\omega}=\sqrt{\omega_c^2 + 4 \omega_o^2}$, $l_e=\sqrt{\hbar/m^*\tilde{\omega}}$,
and $m_r=m^*\tilde{\omega}^2/(4\omega_o^2)$ are respectively, the effective characteristic frequency of
the harmonic oscillator, effective magnetic length, and the renormalized effective mass in the system;
$C=k_l - i (\lambda'+ y''/2)$, $C^+$ is the complex conjugate of $C$, $\lambda'=\lambda_o\,
l_e/\hbar$, $\lambda_o=e E_y/\omega_c$, $\omega_c=e B/\rm{(}m^* c\rm{)}$ is the cyclotron frequency,
$k_l=k\, l_e\, \omega_c/\tilde{\omega}$, $y''=y' \omega_c/\tilde{\omega}$, $y'=y/l_e$, $x'=x/l_e$; and $k$
is the magnitude of the propagation vector, and $\alpha \in \mathbb{R}$ labels the continuous spectrum.
The derivations of Eqs. (2) and (3) are relegated to the Appendix B.

Given the scope and focus, we would like to restrain from expanding on some intricacies regarding, e.g., the
Kohn's theorem [15] and its generalization to quantum wires [16]. It seems vital to underline that, besides
considering a physical system as complete as possible, the present undertaking constitutes a novel proposal
of quantum wires qualifying for the design of nanoscale optical amplifiers -- as compared to the conceptual
tidbits examined in Refs. [17-20].

\subsection{Density-density correlation function}

We start with the general expression of the non-interacting single-particle density-density correlation function
(DDCF) $\chi^0 (...)$ defined as [3]
\begin{equation}
\rchi^{0} ({\bf r},{\bf r'};\omega)=\sum_{ij}\, \Lambda_{ij}\,\,
\psi^*_i ({\bf r})\,\psi_j ({\bf r})\,
\psi^*_j ({\bf r'})\,\psi_i ({\bf r'}),
\end{equation}
where ${\bf r}\equiv (x,y)$, the composite index $i,j\equiv n,\alpha,k$, and $\Lambda_{ij}$ is defined as follows.

\begin{equation}
\Lambda_{ij}= 2\, \frac{f(\epsilon_i)-f(\epsilon_j)}{\epsilon_i-\epsilon_j+\hbar\omega^+},
\end{equation}
where $f(x)$ is the familiar Fermi distribution function. $\omega^*=\omega+i\gamma$ and small but nonzero $\gamma$
represents the adiabatic switching of the Coulomb interactions in the remote past. The factor of $2$ takes care of
the spin degeneracy. Next, we recall the Kubo's correlation function to express the induced particle density [3]
\begin{align}
n_{in}({\bf r},\omega)
=&\int d{\bf r}'\, \rchi({\bf r}, {\bf r}';\omega)\,V_{ex}({\bf r}',\omega)\, , \\
=&\int d{\bf r}'\, \rchi^o({\bf r}, {\bf r}';\omega)\,V_{}({\bf r}',\omega)\, ,
\end{align}
where $V_{tot}=V_{ex}+V_{in}$ is the total potential, with $V_{ex}$ ($V_{in}$) as the external (induced) potential.
$\rchi$ and $\rchi^o$ are, respectively, the total (or interacting) and the single-particle DDCF. Further, the
induced potential in terms of the induced particle density is given by
\begin{equation}
V_{in}({\bf r},\omega)=\int d{\bf r'}\, V_{ee}({\bf r}, {\bf r}')\,n_{in}({\bf r}',\omega)\, ,
\end{equation}
where $V_{ee}(...)$ represents the binary Coulomb interactions and is defined as
\begin{equation}
V_{ee}({\bf r}, {\bf r'})=\frac{e^2}{\epsilon_b}\,\frac{1}{\mid {\bf r}-{\bf r'} \mid}
                         =\frac{e^2}{\epsilon_b}\,\frac{1}{\mid (x-x')^2+(y-y')^2 \mid^{1/2}},
\end{equation}
where $-e$ ($e>0$) is the electronic charge and $\epsilon_b$ the background dielectric constant of the medium, and
its 1D Fourier transform is given by
\begin{equation}
V_{ee}(q; x, x')=\frac{2\,e^2}{\epsilon_b}\,K_o(q\mid x-x' \mid),
\end{equation}
where $K_o(x)$ is the zeroth-order modified Bessel function of the second kind, which diverges as -$\ln (x)$ for
$x\rightarrow 0$. It is quite lengthy but straightforward to demonstrate, from Eqs. (6)$-$(8), that $\rchi$(...)
and $\rchi^o$(...) are, in fact, correlated via the famous Dyson equation [see Fig. 3]

\begin{figure}[htbp]
\includegraphics*[width=8.5cm,height=5cm]{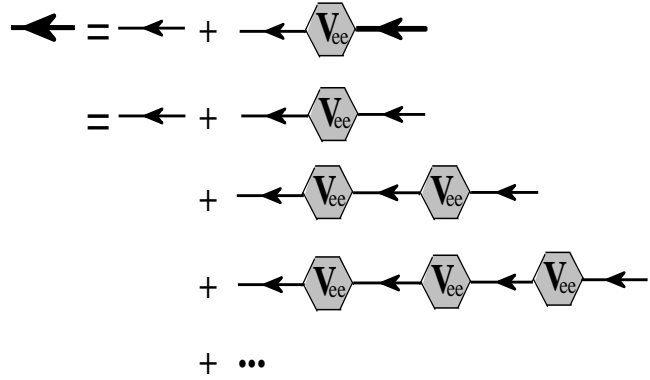}
\centering
\caption{A Feynman diagram for the Dyson equation: the thick [thin] line represents the reducible [irreducible]
DDCF $\rchi (...)$ [$\rchi^o (...)$] in the full RPA. Here $V_{ee}$ represents the binary Coulomb interactions 
in the direct space. The arrows indicate the transition from initial to final spatio-temporal position of the
particle in the process.}
\label{fig3}
\end{figure}
\vspace{-0.5cm}

\begin{equation}
\rchi (x, x') = \rchi^o(x, x') + \int d{x''\,\int d{x'''}\,
                       \rchi^o(x, x'')\, V_{ee}(x'', x''')\,\rchi(x'''}, x'),
\end{equation}
This equation is written such that $\rchi$(...) and $\rchi^o$(...) are both Fourier transformed with respect to 
the spatial coordinate y and ($q, \,\omega$)-dependence is suppressed for the sake of brevity. The Dyson equation
characteristically represents the quantum system and is known to serve useful purpose for investigating various
electronic, optical, and transport phenomena thereof. The Dyson equation, Eq. (11), can be cast in the form
\begin{equation}
\rchi^o(x,x') = \int dx''\,\rchi(x,x'')\,\epsilon(x'',x')\, ,
\end{equation}
where $\epsilon(x,x')$ is the nonlocal, dynamic dielectric function (DF) defined as
\begin{equation}
\epsilon(x,x')=
\delta(x-x') - \int dx''\,\rchi^o(x,x'')\,V_{ee}(x'',x')\, .
\end{equation}
If we multiply both sides of Eq. (12) by the inverse dielectric function
$\epsilon^{-1}(x',x''')$, integrate over $x'$, and make use of the identity:
$\int dx''\,\epsilon(x,x'')\,\epsilon^{-1}(x'',x')=\delta(x-x')$;
we can cast Eq. (12) in the form
\begin{equation}
 \rchi(x, x') = \int dx''\,\rchi^o(x,x'')\,\epsilon^{-1}(x'',x')\, ,
\end{equation}
where $\epsilon^{-1}(x'',x';\omega)$ is the nonlocal, dynamic, inverse dielectric function, which is formally
expressed as follows.
\begin{equation}
\epsilon^{-1}(x,x')=
\delta(x-x') + \int dx''\,V_{ee}(x',x'')\,\rchi(x'',x')\, .
\end{equation}
Both Eqs. (12) and (14) represent the interesting correlation functions, which are essentially made use of in the
studies of the many-particle physics of N-dimensional quantum plasmas. A simple mathematical manipulations of Eqs.
(6)$-$(8), employing the condition of self-consistency: $V_{tot}=V_{ex}+V_{in}$, and imposing the setting of the 
self-sustaining magnetoplasma oscillations ($V_{ex}=0$) yields the generalized, non-local, dynamic dielectric
function to be defined as follows:
\begin{equation}
\epsilon_{nn'mm'}(q;\omega)=\delta_{mn}\delta_{m'n'} - \Pi_{nn'}(q,\omega)\,F_{nn'mm'}(q)\, ,
\end{equation}
where $q=k'-k$ is the momentum transfer, and
\begin{align}
\Pi_{nn'}(q,\omega)&=\frac{1}{L_y}\sum_{k,\alpha} \Lambda_{\stackrel{n\alpha}{n'\alpha'}}(q, \omega)\nonumber\\
                   &=\frac{2}{L_y}\sum_{k,\alpha}
                   \frac{f(\epsilon_{nk\alpha})-f(\epsilon_{n'k'\alpha'})}
                   {\epsilon_{nk\alpha}-\epsilon_{n'k'\alpha'}+\hbar \omega^+}
\end{align}
is, generally termed as the polarizability function, and
\begin{align}
F_{nn'mm'}(q)=\frac{2e^2}{\epsilon_b}\,\int dx\,\int dx'\,
\phi^*_{n\alpha}(x)\,\phi_{n'\alpha'}(x)\,K_o(q\mid x-x'\mid)\phi^*_{m'\alpha'}(x')\,\phi_{m\alpha}(x')\,
\end{align}
refers to the matrix elements of the Coulombic interactions. This implies that the excitation spectrum should
be computed by searching the zeros of det$|\epsilon(q,\,\omega)|$. This also demands limiting the number of
subbands to be involved in the problem; otherwise, the general matrix, with elements $\epsilon_{nn'mm'}$ (...),
is an $\infty\times\infty$ matrix and hence impossible to solve. In order to bring the compact analytical
results obtained hitherto to the practicality, we shall have to adhere to the diagnosis of these results under
certain realizable conditions. This is the task we take up in the next section.

\subsection{Analytical diagnosis}

Ever since the scientists have learnt to grow and control their dimensionality, the quasi-$N$ dimensional (with
$N\le 2$) semiconducting systems have paved the way to some exotic (physical) effects never before seen in the
conventional host materials [3]. The discovery of quantum Hall effects [both integral and fractional] -- that
has changed our basic notions of how an external magnetic field at temperatures close to zero can pave the way
to some  unprecedented quantum states -- is one of them. This tells us why the succeeding experiments on the
systems of reduced dimensionality have generally been performed at lower temperatures. In order to turn the
greater number of our analytical results into practicality, we intend to diagnose them at zero temperature,
with limited number of occupied subbands, and simplified by the symmetry of the confining potential. The zero
temperature limit is of utmost practical interest because the vast majority of experiments employed on these
systems to observe electronic, optical, and transport phenomena have been performed at close to absolute zero.

\subsubsection{Limiting the number of subbands}

The generalized dielectric function matrix represented by Eq. (16) is, in general, an $\infty\times\infty$ matrix
until and unless we delimit the number of subbands (and hence limit the electronic transitions) involved in the
problem. Also, it should be pointed out that while experiments may report multiple subbands occupied, theoretically
it is extremely difficult to compute the excitation spectrum for the multiple-subband model. The reason is that the 
generalized dielectric function turns out to be a matrix of the dimension of $\eta^2\times\eta^2$, where $\eta$ is
the number of subbands accounted for. Handling such enormous matrices (for a very large $\eta$) analytically is a
hard nut to crack and then no remarkable fundamental science (or technology) is ever expected to be emerging out of
such undue complexity. For this reason, we choose to keep the complexity to a minimum and confine ourselves to a
two-subband model ($n,\,n',\,m,\,m'\equiv$ 1, 2) with only the lowest one occupied. This is a very reasonable
assumption for these low-density, low-dimensional systems at lower temperatures where most of the experiments are
carried out. This implies that the generalized dielectric function becomes a $4\times4$ matrix define by
\begin{equation}
\tilde{\epsilon}(q,\omega)=
\left [
\begin{array}{rrrr}
1-\Pi_{11}\,F_{1111} \ & \ -\Pi_{11}\,F_{1112} \ & \ -\Pi_{11}\,F_{1121} \ & \ -\Pi_{11}\,F_{1122} \\
-\Pi_{12}\,F_{1211} \ & \ 1-\Pi_{12}\,F_{1212} \ & \ -\Pi_{12}\,F_{1221} \ & \ -\Pi_{12}\,F_{1222} \\
-\Pi_{21}\,F_{2111} \ & \ -\Pi_{21}\,F_{2112} \ & \ 1-\Pi_{21}\,F_{2121} \ & \ -\Pi_{21}\,F_{2122} \\
-\Pi_{22}\,F_{2211} \ & \ -\Pi_{22}\,F_{2212} \ & \ -\Pi_{22}\,F_{2221} \ & \ 1-\Pi_{22}\,F_{2222} \\
\end{array}
\right ].
\end{equation}
Here we need to underline that $\Pi_{22} = 0$, since the second subband is assumed to be unoccupied. Since the
quasi-particle excitations are determined by searching the zeros of det\,$|\epsilon(q,\,\omega)|$, Eq. (19)
finally yields
\begin{align}
\Big(1 - \Pi_{11} F_{1111}\Big) \Big(1 - \rchi_{12} F_{1212}\Big) - \Pi_{11} \rchi_{12} F^2_{1112} = 0 ,
\end{align}
where $\rchi_{12} =\Pi_{12} + \Pi_{21}$ is the intersubband polarizability function that takes both the upward and the 
downward transitions into account. This is, in fact, the final analytical result to be treated at the computational
level in order to obtain the single-particle as well as the collective (magnetoplasmon) excitation spectrum in the
Q-1DEG system at hand. Notice, however, that further simplification may be warranted depending upon the nature of the
confining potential in the problem (see what follows in the next section).

\subsubsection{Symmetry of the confining potential}

One of the intriguing details of the numerical computation of the collective excitations in any system is that for a
symmetric confining potential $F_{ijkl}$($q$) -- the matrix element of the Fourier-transformed Coulomb interaction --
is strictly zero for arbitrary value of $q$ provided that the sum of the subband indices ($i+j+k+l$) is an odd number.
This is because the corresponding wave function is either symmetric or antisymmetric under space reflection. But this
is the case that prevails only in the absence of an applied magnetic field. In the presence of an applied magnetic
field (as is the case here), the situation takes a different turn. In the case of $B\ne 0$, even though the confining 
potential along the x direction is symmetric and the sum $i+j+k+l$ is an odd number, $F_{ijkl}$($q$) does {\it not}
apparently seem to be stringently zero. This is attributed to the fact that in the presence of a finite magnetic
field the center of the cyclotron orbit is displaced from zero by a finite distance $x_c$ [$=k_{\ell}$]. However, this
is not truly the endgame of the illusionary analytical results for $F_{ijkl}$($q$) in the case of finite $B$. One can 
carefully play with the integrands using the two-stage substitutions to prove that $F_{ijkl}$($q$) is indeed zero
even for the $B\ne0$ if the sum $i+j+k+l=$ odd. An immediate consequence of this is that the coupling term $F_{1112}$
($q$) in Eq. (20) becomes zero and hence the intrasubband and intersubband magnetoplasmon excitations become decoupled.
Since the hybrid subband index $n$ is allowed the value 0 (1) for the intrasubband (intersubband) excitations, we
still need to conform Eq. (20) such that the subscript $1 \rightarrow 0$ and $2 \rightarrow 1$ for all practical
purposes. The ready-to-feed expressions determined for $F_{0000}$($q$), $F_{0101}$($q$), and $F_{0001}$($q$)
[see Eq. (18)] are given by
\begin{equation}
F_{0000}(q)=
\frac{1}{\pi}\,\frac{2 e^2}{\epsilon_b}\,Z\,\int dy\,\int dy'\, e^{-y^2}\, K_0\big(\sqrt{2}q\ell_e\mid y-y'\mid\big)
\, e^{-y'^{2}}\, ,
\end{equation}
\begin{equation}
F_{0101}(q)=
\frac{1}{\pi}\,\frac{2 e^2}{\epsilon_b}\,Z\,\int dy\,\int dy'\,e^{-y^2}\, y\,K_0\big(\sqrt{2}q\ell_e\mid y-y'\mid\big)
\, y'\, e^{-y'^{2}}\, ,
\end{equation}
and
\begin{equation}
F_{0001}(q)=
\frac{1}{\pi}\,\frac{2 e^2}{\epsilon_b}\,Z\,\int dy\,\int dy'\,e^{-y^2}\, K_0\big(\sqrt{2}q\ell_e\mid y-y'\mid\big)
\, y'\, e^{-y'^{2}}\, ,
\end{equation}
where $Z=\exp\{-2\,y^2_o\}$, $y_o=q_{\ell}/\sqrt{2}$, and $q_{\ell}=q\,\ell_e\,(\omega_c/\tilde{\omega})$. Note that
$y$ and $y'$ are the dimensionless variables. The computation of $F_{0000}$($q$), $F_{0101}$($q$), and $F_{0001}$($q$)
clearly substantiates the foregoing notion that $F_{0001}$($q$)$=0$ for an arbitrary value of $q$. It is observed
that $F_{0000}$($q$) turns out to be predominant over the entire range of wave vector $q$. This is clearly due to the
fact that the charge density at lower temperatures is largely concentrated in the ground state.

\subsubsection{The zero-temperature limit}

Ever since the systems of reduced dimensionality gained the momentum, it has become widely known that most of the
experiments on these quantal structures are performed at extremely low temperatures. Therefore, we choose to confine
ourselves to the absolute zero temperature. Limiting to absolute zero has certain edge over the finite temperatures:
(i) this allows us to replace the Fermi distribution function with the Heaviside step function such that
\begin{equation}
f(\epsilon)=\theta(\epsilon_F - \epsilon)
=\left \{
\begin{array}{c}
1 \, , \,\,\,\, {\rm if \,\,\,\,\, \epsilon_F > \epsilon}\\
0 \, , \,\,\,\, {\rm if \,\,\,\,\, \epsilon_F < \epsilon}
\end{array}
\right. \, ,
\end{equation}
where $\epsilon_F$ is the Fermi energy in the system, (ii) this helps us convert the sum over ${k}$ to an integral by
using the summation replacement convention with respect to the 1D such that
\begin{align}
\sum_{k} \rightarrow \frac{L_y}{(2\pi)}\,\int^{{k}_F}_{-{k}_F} d{k}\, ,
\end{align}
and, most importantly, (iii) this permits us to calculate analytically the manageable forms of polarizability function $\Pi_{nn'}(...)$ to write, e.g.,
\begin{align}
\Pi_{_{00}}(q, \omega)=\frac{m_r}{\pi q \hbar^2}\,
\ln \left[\frac{(\hbar \omega)^2 - (\epsilon_q -\epsilon_{\lambda} - \hbar q v_F)^2}
               {(\hbar \omega)^2 - (\epsilon_q -\epsilon_{\lambda} + \hbar q v_F)^2}\right]
\end{align}
and
\begin{align}
\rchi_{_{01}}(q, \omega)=\frac{m_r}{\pi q \hbar^2}\,
\ln \left[\frac{(\hbar \omega)^2 - (\epsilon_q -\epsilon_{\lambda} + \Delta_{01} - \hbar q v_F)^2}
               {(\hbar \omega)^2 - (\epsilon_q -\epsilon_{\lambda} + \Delta_{01} + \hbar q v_F)^2}\right]
\end{align}
where $\epsilon_q = \hbar^2 q^2/2 m_r$, $\epsilon_{\lambda}=\lambda \hbar\beta/m^*$, $\beta=(\alpha'-\alpha)$, 
$\Delta_{01}=\hbar \tilde{\omega}$, and $v_F=\hbar k_F/m_r$. Here $k_F$ and $v_F$ are, respectively, the Fermi wave
vector and Fermi velocity in the system. In the long wavelength limit (i.e., $q \to 0$), Eqs. (26) and (27) -- after
lengthy and sagacious mathematical manipulation -- yield
\begin{equation}
\Pi_{_{00}}(q, \omega)=\frac{n_{_{1D}}\, q^2 }{m_r \omega^2}\big(1-\epsilon_{\lambda}/\epsilon_{q}\big) + O(q^4),
\end{equation}
and
\begin{equation}
\rchi_{_{01}}(q, \omega)=n_{_{1D}}\, \frac{2\,(\Delta_{01}-\epsilon_{\lambda})}
                                     {\big[(\hbar \omega)^2-(\Delta_{01}-\epsilon_{\lambda})^2\big]} + O(q),
\end{equation}
Comparing $\Pi_{_{00}}$ with $\rchi_{_{01}}$ leads us to infer that both $\Pi_{_{00}}$ and $\rchi_{_{01}}$ are directly proportional to the 1D charge density $n_{_{1D}}$. In addition, $\Pi_{_{00}} \propto q^2$ if
$\epsilon_{\lambda}/\epsilon_q \ll 1$ and $\rchi_{_{01}} \propto \Delta_{01}$ if $\Delta_{01} \gg \epsilon_{\lambda}$.

\subsubsection{The DOS and the Fermi energy}

In condensed matter physics, (electronic) density of states (DOS) is one of the central importance because it defines
and dictates the behavior of a quantal system and is fundamental to the understanding of its electronic, optical, and 
transport phenomena. Equally significant is the notion of the Fermi energy because all the transport phenomena of a
fermionic system are governed by the electron dynamics at (or near) the Fermi surface. In the standard physical
conditions, the DOS and the Fermi surface influence each other in quite an intricate fashion. Here we are interested
in studying the DOS and the Fermi energy in a system of quantum wires in the presence of a longitudinal electric
field and an applied magnetic field in the symmetric gauge. We start with Eq. (3) to derive the following expression
\begin{equation}
D(\epsilon)=\frac{1}{\pi}\,\Big(\frac{2 m_r}{\hbar^2}\Big)^{1/2}\,
\sum_{n}\, \big[\epsilon - \epsilon_{n\lambda} \big]^{-1/2}\, \theta (\epsilon - \epsilon_{n\lambda})\, ,
\end{equation}
for computing the density of states and
\begin{equation}
n_{1D}=\frac{2}{\pi}\,\Big (\frac{2 m_r}{\hbar^2}\Big )^{1/2}\,
\sum_{n}\, \big[\epsilon_F - \epsilon_{n\lambda}\big]^{1/2}\, \theta (\epsilon_F - \epsilon_{n\lambda})
\end{equation}
for computing the Fermi energy. Notice that Eq. (31) is to be treated as a transcendental function and solved 
self-consistently with two loops on $n_{1D}$ and $\epsilon_F$ running simultaneously. The symbol $\epsilon_{n\lambda}=\epsilon_n-\epsilon_{\lambda}-\lambda^2_o/(2m^*)$; with
$\epsilon_n=(n+1/2)\,\hbar \tilde{\omega}$ and $\epsilon_{\lambda}=\lambda_o \hbar\alpha/m^*$.
Here $k_F=(\pi/2)\,n_{1D}$ is the Fermi wave vector, and $n_{1D}$
is the linear density (i.e., the number of electrons per unit length) of the system. Typically, it is customary to
subtract the zero-point energy $\epsilon_0$ ($=\frac{1}{2}\hbar\tilde{\omega}$) from the Fermi energy to compute the
effective Fermi energy of a system.

\subsubsection{The optical gain coefficient}

The crux of the problem undertaken in the present work is the evolution of an intersubband magnetoplasmon excitation
into a magnetoroton (MR) above a threshold value of an applied magnetic field. The MR exhibits a negative group
velocity (NGV) between the maxon and the roton. The manifestation of NGV implies anomalous dispersion in a
gain medium -- with the population inversion -- in which the stimulated emission causes amplification of the
incoming light of a certain wavelength. In other words, the gain medium with population inversion forms the basis
for the lasing action of lasers. We are motivated to determine the gain coefficient $\alpha (\omega)$ in the system
of quantum wires under the given physical conditions.

The science behind the lasing technology researched largely after the sixties reveals that there is no specific gain
mechanism that is of fundamental importance in the subject. This is because the classic Kramers-Kronig (KK)
relations, which allow us to express the real (imaginary) part in terms of imaginary (real) part of any complex
response function that is analytic in the upper half-plane, produce identical dispersive effects regardless of the
origin of the gain. In the present case, the complex response function is obviously the total (or interacting) DDCF
$\rchi (...)$ [see, e.g., Eq. (11) or (14)]. The imaginary part of $\rchi (q, \omega)$ represents the gain or loss in
the system, as the case may be [21], which we make use of in defining the gain coefficient
\begin{align}
\alpha (\omega)=\hbar c\,\frac{1}{\pi}\,
\left\{
 P\int^{\infty}_{-\infty}\! d\omega'\,\,\frac{\omega'\,\rm{Im}[\rchi (q, \omega'; q_x)]}{{\omega'}^2-\omega^2}
+P\int^{\infty}_{-\infty}\! d\omega'\, \,\frac{\omega\,\rm{Im}[\rchi (q, \omega'; q_x)]}{{\omega'}^2-\omega^2}
\right\}\, ,
\end{align}
where $P$ denotes the Cauchy principal value and the universal constants -- $\hbar$ and $c$ -- have been introduced
merely to keep the $\alpha (\omega)$ dimensionless. Since $\rchi (\omega')$ is an odd function, the second integral
vanishes, and we are left with
\begin{align}
\alpha (\omega)=\hbar c\,\frac{2}{\pi}\,
 P\int^{\infty}_{0}\! d\omega'\,\,\frac{\omega'\,\rm{Im}[\rchi (q, \omega'; q_x)]}{{\omega'}^2-\omega^2}\, .
\end{align}
This definition of the gain coefficient $\alpha (\omega)$ is quantal analogue of the Lorentz-Drude model employed
in classical electromagnetism. We compute $\alpha (\omega)$ for a given value of the propagation vector $q$ and
for $q_x=0$ -- until and unless stated otherwise. It is noteworthy that both Eqs. (11) and (14) -- while seem to be
apparently different -- produce exactly identical results. As to the inverse dielectric function (IDF) in Eq. (14),
Ref. 21 offers the exact analytical results for evaluating the IDF. The numerical results are discussed in the next
section. While the present proposal exploits the notion of lasing with (the population) inversion, the idea of
lasing without inversion (LWI) [22] has also been gaining ground in the recent past [23].

\vspace{-0.5cm}

\section{Illustrative Examples}

As to the specific illustrative examples, we focus on the narrow channels of the In$_{1-x}$Ga$_x$As system with the
effective mass $m^*=0.042 m_o$ and the background dielectric constant $\epsilon_b=13.9$. We compute the collective 
(magnetoplasmon) excitations in a Q-1DES for a two-subband model within the framework of random-phase approximation
(RPA) at T=0 K (since extremely low temperatures are preferred for most experiments on these systems) [3]. This is
done by examining the impact of several parameters involved in the analysis such as the 1D charge density $n_{1D}$, 
characteristic frequency of the harmonic potential $\omega_o$, magnetic field $B$, and the electric field $E_y$.
The effective confinement width of the harmonic potential, estimated from the extent of eigenfunctions, is
$w_{eff}=28.69$ nm for $B=3.0$ T and $\hbar\omega_o=1.53$ meV. Notice that the Fermi energy $\epsilon_F$ varies with
$n_{1D}$, $\omega_o$, $B$, or $E_y$.

\begin{figure}[htbp]
\includegraphics*[width=8.5cm,height=8.5cm]{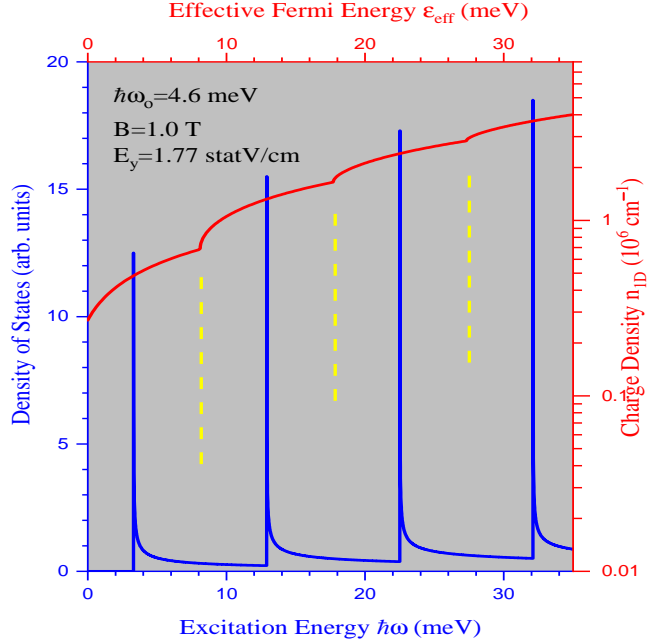}
\centering
\caption{(Color Online) The density of states versus the excitation energy (in Blue) and the effective Fermi
energy versus the charge density (in red). The broken (yellow) lines (only) indicate the dips (in the Fermi
energy) lying exactly in the center of the nearest DOS peaks.}
\label{fig4}
\end{figure}

The literature is live witness for quite some time that the density of states (DOS) for the quantum wires depends
on the energy [with a power of −(1/2)] and that it takes on the spikes separated by the magnitude of the
{\it effective} confinement potential [3]. However, it does not seem to have been pointed out before that the
Fermi energy shows typical (periodic) dips, which lie exactly midway between the consecutive peaks of the DOS. It
is observed that the larger the ({\it effective}) confinement potential ($\hbar\tilde{\omega}$), the smaller the
number of such spikes (and hence of dips). This is demonstrated in Fig. 4, which plots both the DOS and the Fermi
energy on different scales for specific values of the charge-density and the confinement potential.

\begin{figure}[htbp]
\includegraphics*[width=9.5cm,height=8.5cm]{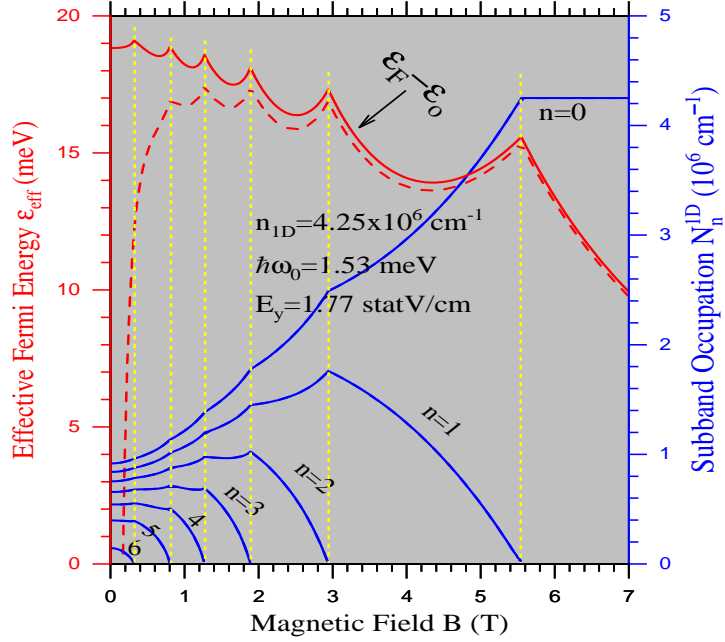}
\centering
\caption{(Color Online) The variation of the Fermi energy $\epsilon_F$ (relative to the ground-state sublevel
$\epsilon_0$) and subband occupation $N^{1D}_n$ with the applied magnetic field $B$. Notice the attestation
of the broken (yellow) vertical lines.}
\label{fig5}
\end{figure}

Figure 5 shows the variation of the effective Fermi energy $\epsilon_{eff}$ and the subband occupation $N^{1D}_n$
with $B$. Given the complexity of the electronic spectrum, one cannot expect the
Fermi energy to be a smooth function of $B$. The solid (dashed) curves represent the case of $E_y=0$ ($E_y \ne 0$).
It is evident that there are {\it seven} occupied subbands before the magnetic field is switched on. The reader is
reminded that the consecutive peaks in the density of states are seen to be separated exactly by the subband
separation $\hbar\tilde{\omega}$ and this separation increases with increasing $B$. Due to this diamagnetic shift,
a magnetic depopulation of the subbands occurs and oscillations akin to SdH oscillations appear in the
magnetoresistance. The role of electric field is fundamental: it remains indiscernible for the individual subbands
but becomes noticeable for the Fermi energy -- the reason being the summation over the subband index $n$ for the
latter. The (yellow) vertical lines testify how the peaks in the Fermi energy, the tips in the subband occupation,
and the point of total depopulation of the next subband concur at the same value of the magnetic field $B$.

\begin{figure}[htbp]
\includegraphics*[width=8.5cm,height=8.1cm]{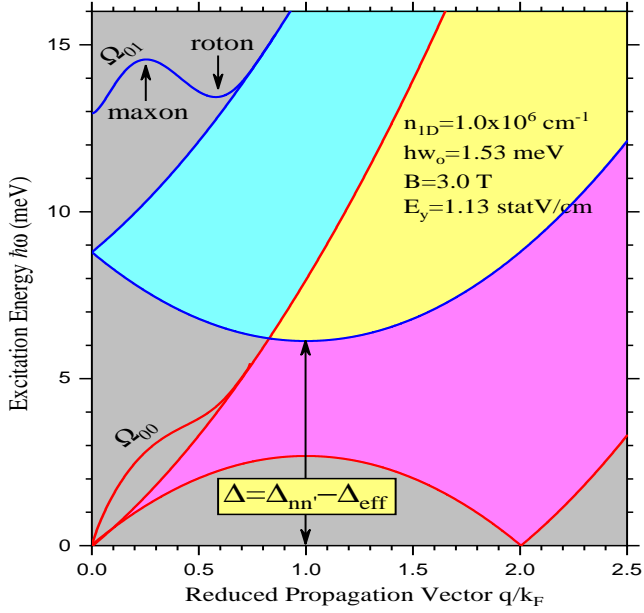}
\centering
\caption{(Color Online) The full magnetoplasmon spectrum plotted as energy $\hbar\omega$ vs. reduced wave vector
$q/k_F$. The magenta (cyan) shaded region refers to the intrasubband (intersubband) SPE -- with {\it shared}
yellow shaded region. The bold curve marked $\Omega_{00}$ ($\Omega_{01}$) stands for the intrasubband
(intersubband) CME. The vertical arrow measures the minimum of intersubband SPE at $q=k_F$. The MR -- marked
$\Omega_{01}$ -- is the centerpiece of this article.}
\label{fig6}
\end{figure}

Figure 6 illustrates the full magnetoplasmon excitation spectrum plotted as the energy $\hbar\omega$ versus the
reduced wave vector $q/k_F$, for the given values of $n_{1D}$, $\hbar\omega_o$, $B$, and $E_y$. The figure
caption specifies both the single-particle excitations (SPE) and the collective (magnetoplasmon) excitations
(CME) in the system. The intrasubband CME starts from the origin and merges with the upper edge of the
intrasubband SPE at ($q/k_F=0.735$, $\hbar\omega=7.345$ meV). The intersubband CME starts at ($q/k_F =0$,
$\hbar\omega=12.946$ meV), attains a maximum at ($q/k_F =0.255$, $\hbar\omega=14.559$ meV), reaches its minimum
at ($q/k_F =0.579$, $\hbar\omega=13.433$ meV), and then rises up to merge with the upper edge of the intersubband
SPE at ($q/k_F =0.856$, $\hbar\omega=15.291$ meV). After merging with their respective SPE, both CME become
Landau damped. It is not difficult to check (analytically) the single-particle energies at the critical points
such as $q/k_F=0$, 1, and 2. The most interesting aspect of the excitation spectrum is the intersubband CME
(referred to as the MR). The MR acquires a negative group velocity (NGV) between the maxon and the roton. At the
origin, the energy difference between the SPE and the MR is a manifestation of the many-body effects such as the 
depolarization and excitonic shifts [3].

\begin{figure}[htbp]
\includegraphics*[width=8.5cm,height=8.25cm]{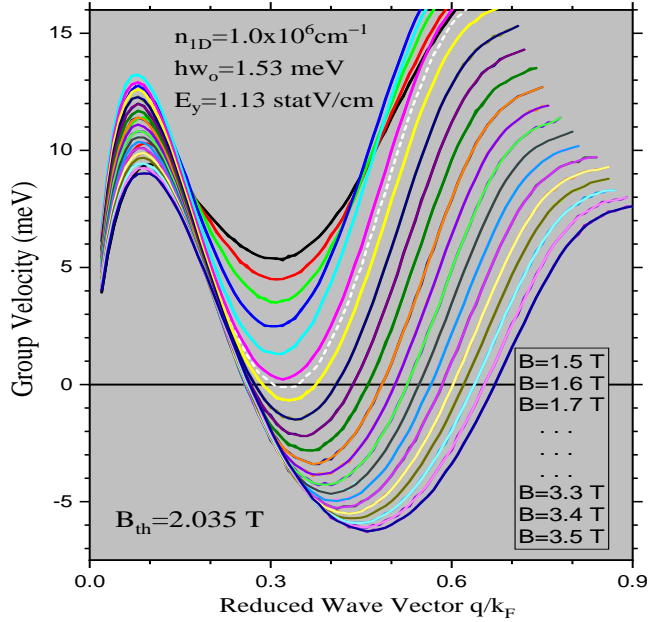}
\centering
\caption{(Color Online) The group velocity of the MR versus the wave vector $q/k_F$ for several values of $B$ --
and for the given $n_{1D}$, $\hbar\omega_o$, and $E_y$. The white dashed curve stands for the threshold value of
$B$ ($B_{th}=2.035$ T).}
\label{fig7}
\end{figure}

Figure 7 exhibits the group velocity of the MR (see Fig. 1) as a function of the propagation vector $q/k_F$, for
several values of $B$. The group velocity ($V_g$) is measured in meV because we computed $V_g$ specified by
$V_g = d(\hbar\omega)/d Q$, where $Q = q/k_F$. The ICE is seen to attain its MR character only for $B \ge B_{th}$
where the $V_g$ curves cross the zero twice: first for the maxon and second for the roton.
Figure 7 establishes the fact that the MR does secure an NGV between the maxon and the roton. Since the occurrence
of NGV is not quite recurrent, there must be some dramatic consequences. It turns out that the NGV does matter in
the phenomena such as tachyon-like (superluminal) behavior [24-26], anomalous dispersion in the gain medium, a state
with population inversion (typically) characterized by the negative temperature, ...etc.

\begin{figure}[htbp]
\includegraphics*[width=8.5cm,height=8.5cm]{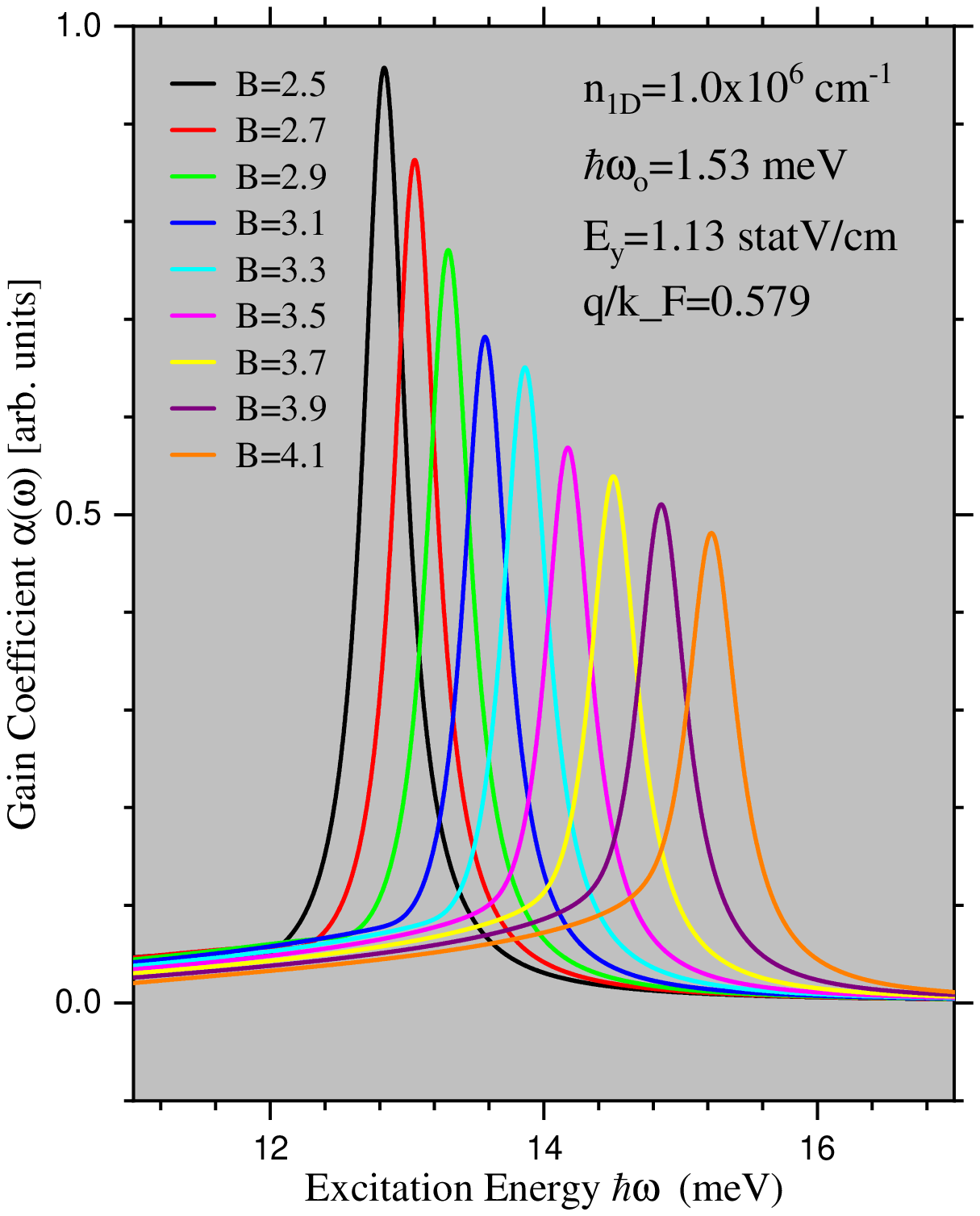}
\centering
\caption{(Color Online) The gain coefficient $\alpha(\omega)$ versus the excitation energy $\hbar\omega$
for several values of the magnetic field $B$ -- and for the given values of $n_{1D}$, $\hbar\omega_o$,
and $E_y$.}
\label{fig8}
\end{figure}

Figure 8 exemplifies the computation of gain coefficient $\alpha(\omega)$ so as to visualize the notion of a
quantum wire acting as an optical amplifier. The gain coefficient experiences a blue shift with the increasing
$B$. It is interesting to notice that while the amplitude of $\alpha(\omega)$ decreases with increasing $B$,
the bandwidth of the laser amplifier remains practically unaltered. The notion of {\it bandwidth} in laser
amplification is different from that in the band structures. The bandwidth of an amplifier is defined as the
full distance between the energy points at which the gain has dropped to half the peak value. Another critical
issue is the nature of electronic transitions: an amplifying (absorbing) transition implies a positive (negative)
values of $\alpha(\omega)$. Obviously, one can always assign a suitable $\pm$ sign with $\alpha(\omega)$
to give it a proper meaning -- although it is the calculus that should dictate the sign convention. These remarks
on the sign convention are fully supported by the literature on lasers.

The existence of an MR excitation in the system brings on the notion that the applied magnetic field drives the
system to a metastable (or non-equilibrium) state. The metastable state is an immensely significant concept in
condensed matter physics. It is defined to be a state that {\it may} exist albeit it is much less stable than
its final (equilibrium) state. As we get it, the irradiation of the system with the light of suitable
wavelength allows its electrons to jump to an excited state. When the inciting radiation is removed, the excited
electron tends to fall back to its original (lower) state. However, when an electron goes to a metastable state,
it stays there for a relatively longer duration. This process causes electrons to accumulate in the metastable
state, since their rate of excitation is greater than that of de-excitation. This leads to the occurrence
called {\it population inversion} that forms the basis of the lasing action of lasers.

\begin{figure}[htbp]
\includegraphics*[width=8.5cm,height=8.5cm]{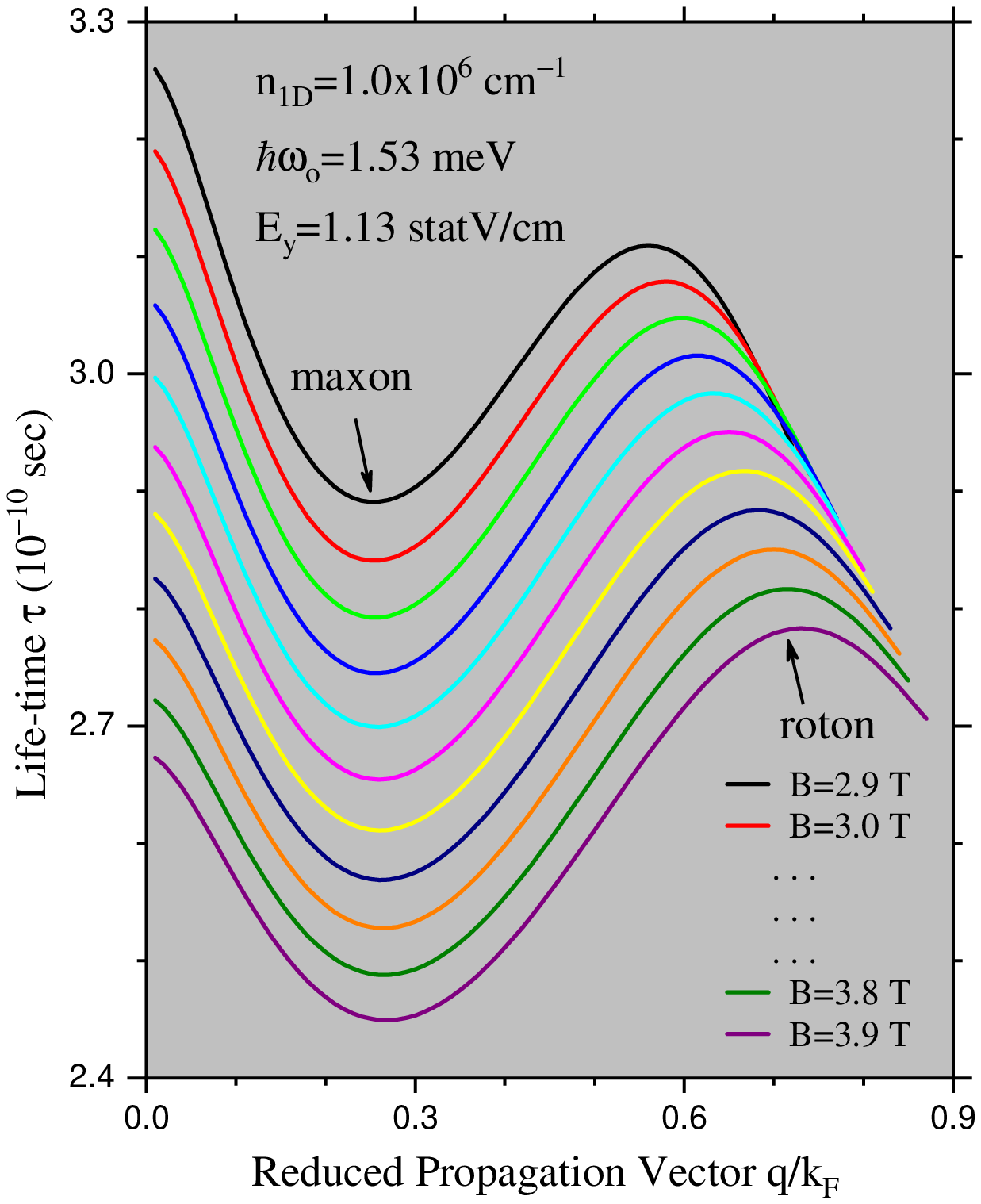}
\centering
\caption{(Color Online) The life-time $\tau$ versus the propagation vector $q/k_F$ for the MR for several values
of $B$ ($\ge B_{th}$). Notice the trend: the stronger the $B$, the shorter the $\tau$, and more susceptible the
metastable state.}
\label{fig9}
\end{figure}

There are various ways to represent the metastable state of a system. We choose to compute the life-time of the
MR in the $\omega$-$q$ space (Fig. 9) -- which is one of the significant means to represent the robustness of
the metastable state in the system  -- until it ceases to exist as a bonafide CME [27]. Just as expected, the MR
sits as an unstable transition state in close proximity of the roton minimum. The picture speaks for itself: the
stronger the magnetic field, the shorter the life-time, and more susceptible the metastable state. This implies
that the moderate magnetic fields favor the optimum case of higher gain.

\begin{figure}[htbp]
\includegraphics*[width=8.5cm,height=9.0cm]{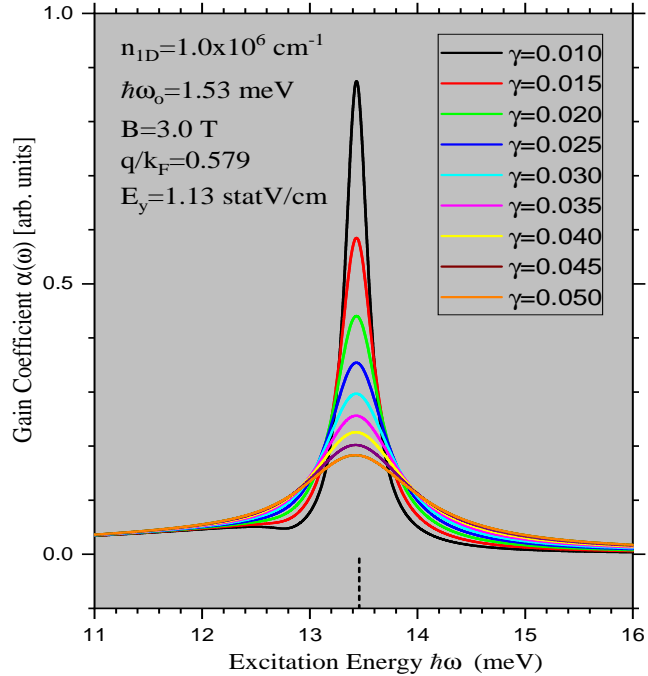}
\centering
\caption{(Color online) The gain coefficient $\alpha (\omega)$ as a function of the excitation energy $\hbar\omega$
for several values of the damping factor and for the given values of the magnetic field $B$, the charge density
$n_{1d}$, and the confinement potential energy $\hbar\omega_o$. The parameters are as given inside the picture.}
\label{fig10}
\end{figure}

Figure 10 illustrates the computation of the gain coefficient $\alpha(\omega)$ as a function of the excitation
energy $\hbar\omega$ for various values of the damping factor $\gamma$. The gain coefficient in the context of
the laser amplification is computed only in the frequency range of the relevant magnetoroton excitations. The
gain coefficient that persists due to the electronic transitions shows a maximum at $\hbar\omega \simeq 13.434$
meV for the damping factor $\gamma=0.010$. The peak position legitimately occurs at the expected values of
($q, \omega$) in the excitation spectrum. As per intuitive conviction, the gain peak turns towards the lower
energy with increasing damping factor. It is reasonable that an amplifier device such as a laser gain medium
cannot maintain a fixed gain for arbitrarily high input powers, because this would demand adding arbitrary
amounts of power to the amplified signal. Therefore, the gain must be reduced for high input powers: this
phenomenon is called {\it gain saturation}. In the case of a laser gain medium, it is widely known that the gain
does not instantly adjust to the level according to the optical input power, because the gain medium stores some
amount of energy, and the stored energy determines the gain.

\section{Concluding Remarks}

In summary, crucial to the motivation behind the present investigation is the NGV between the maxon and the roton.
A typical feature of NGV is that it leads to a superluminal behavior without one's having to introduce negative
energies. The NGV is associated with the anomalous dispersion in a gain medium with
inverted population -- a clear result of the sustained metastable state -- so that gain instead of loss occurs at
the frequencies of interest. A medium with population inversion has the remarkable ability of amplifying a small
optical signal of definite wavelength.

The roton minimum is the mode of higher density of states and its features are among the most significant
manifestations of many-particle interactions. They emerge from an interplay between the direct and exchange terms
of the electron gas and the depth of the minimum is determined by the strength of the exchange-vertex corrections.
As such, incorporating the many-body effects should give a better sense of the texture of MR. Generalizing the
current approach to include third and fourth subbands might also give a deeper insight into the subject. This
fundamental investigation suggests an interesting application: the quantum wires can enable us to design
{\it magnetorotonic optical amplifiers} (at nanoscale with minimal dissipation loss) and hence pave the way to a
new generation of lasers. Since all the parameters -- such as the charge density, confining potential, magnetic
field, and electric field (see, e.g., Fig. 6) -- involved in the process behind this proposal are within the reach
of the current technology, the core concept should prompt the device experimentation and hence turn this rationale
into reality.
\begin{acknowledgments}
It is a pleasure for me to thank Hiroyuki Sakaki, Bahram Djafari-Rouhani, Allan MacDonald, Peter Nordlander,
and Douglas Natelson for the support, discussions, and encouragement. I would also like to thank Kevin Singh
for the timely assistance with the software.
\end{acknowledgments}
\newpage

\appendix

\section{On the gauge invariance}

Here we want to capture briefly the essence of the gauge invariance. It is significant since the Hamiltonian
lacks the translational invariance along the y direction due to the symmetric gauge and causes concern: How
is this compatible with a propagation vector {\bf k} along y? We defied the norm for the following reason.
Let us first recall the standard conventions: Both the Landau gauge (LG) ${\bf A} = B (-y, 0)$ or ${\bf A} =
B (0, x)$ and the symmetric gauge (SG) ${\bf A} = (B/2) (-y, x)$ correspond to ${\bf B} = B {\hat {\bf z}}$.
The former choice of the LG preserves the translational invariance along the x-axis, while the latter
preserves that along the y-axis. The SG preserves the rotational invariance. That means the magnetic field
${\bf B}$ is gauge invariant (and so is the physics of a given system). This also implies that the system is
both translationally and rotationally invariant around the z axis. However, the choice of ${\bf A}$ is not!
The gauge invariance is brought about by the gauge transformation, which is nothing other than changing the
vector potential ${\bf A}$ in such a way that the electromagnetic fields ${\bf E}$ and ${\bf B}$ remain the
same. Suppose we want to go from ${\bf A}_1$ to ${\bf A}_2$. To maintain the same ${\bf E}$ and ${\bf B}$, the
two must differ by at most a gradient such as, e.g., ${\bf A}_2 (x) = {\bf A}_1 (x) + {\bf \nabla} \rchi (x)$.
The net result is that the eigenfunctions transform by a space-dependent phase factor, while the eigenenergies
remain gauge invariant.

\section{On the derivation of eigenfunction and eigenenergy}

Given predominantly the free propagation along the y direction, Eq. (1) can be cast in the following form:
\begin{align}
H=\frac{1}{2}\,\hbar\tilde{\omega}\,
\Big[\Big(\frac{x'}{2}-i\partial_{y''}\Big)^2 + \Big(\frac{y''}{2}+i\partial_{x'}\Big)^2\Big]
+\frac{\hbar^2\,k^2}{2\,m_r} + \lambda\,y''
\end{align}
where $x'=x/\ell_e$, $y'=y/\ell_e$, $y''=(\omega_c/\tilde{\omega})y'$, $\partial_{x'}=\ell_e \partial_x$, $\partial_{y'}=\ell_e \partial_y$, $\partial_{y''}=(\omega_c/\tilde{\omega})\partial_{y'}$, and $\lambda=e\,E_y\,\ell_e\,\tilde{\omega}/\omega_c$ -- with $\partial_r=\partial/\partial r$.
Here $\ell_e=\sqrt{\hbar/m^*\tilde{\omega}}$, $\tilde{\omega}=\sqrt{\omega_c^2+4\omega_o^2}$, and $m_r=m^* (\tilde{\omega}/2\omega_o)^2$
are, respectively, the effective magnetic length, the effective characteristic frequency of the quantum oscillator,
and the renormalized effective mass in the problem. The next step is to introduce the so-called ladder operators,
which provide a convenient means to extract the eigenenergies without directly solving the system's differential
equations. They are generally made use of in the formalism of quantum harmonic oscillator and angular momentum.
It is imperative to recall that the term ladder operator is typically used to describe an operator that acts to
increment or decrement a quantum number describing the state of a system. To change the state of a particle -- in
quantum field theory, for example -- requires the use of an annihilation (creation) operator to remove (add) a
particle from the initial (to the final) state. This clearly dispels the confusion regarding the relationship
between the ladder operators in linear algebra and the creation/annihilation operators commonly used in quantum
field theory. Another simpler way to state is that the ladder operator is an operator that increases or decreases
the eigenvalue of another operator in the problem. As such, we define the ladder operators
\begin{align}
a   =  \frac{1}{\sqrt{2}}\big[z^+ + p^{+}_z - i\lambda' \big], \,\,\,\,\,\,\,\,\,
a^+ =  \frac{1}{\sqrt{2}}\big[z   -   p_z     + i\lambda' \big],
\end{align}
and
\begin{align}
d   =  \frac{-i}{\sqrt{2}}\big[z^+ -  p^{+}_z \big], \,\,\,\,\,\,\,\,\,
d^+ =  \frac{i}{\sqrt{2}}\big[z +  p_z \big],
\end{align}
where
\begin{align}
\begin{array}{c}
z   =  \frac{1}{2}(x' + i y''),\,\,\,\,\,\,\,\,\, p_z = (\partial_{x'} + i \partial_{y''})\\
z^+ =  \frac{1}{2}(x' - i y''),\,\,\,\,\,\,\,\,\,p^+_z= (\partial_{x'} - i \partial_{y''})
\end{array}
\end{align}
and $\lambda'=\lambda/(\hbar \tilde{\omega})$. These operators follow certain commutation relations such as for
example:
\begin{align}
[a^+, a] =1 = [d^+, d]
\end{align}
Thus the Hamiltonian in Equation (B1) assumes the form:
\begin{align}
H = \Big(a^+ a +\frac{1}{2}\Big)\, \hbar\tilde{\omega} - \frac{\lambda}{\sqrt{2}}\big(d+d^+\big) +
    \frac{\hbar^2\,k^2}{2\,m_r} - \frac{1}{2} \lambda \lambda'
\end{align}
In order to determine the eigenfunction $\psi$ and eigenenergy $E$, we split Eq. (B6) into two mutually commuting
parts as follows:
\begin{align}
H=H_{osc} - H_{lin},
\end{align}
where $H_{osc}$ ($H_{lin}$) represents the oscillatory (linear) part of the total Hamiltonian. To be explicit,
the oscillatory part $H_{osc}$ is given by
\begin{align}
H_{osc} = \big(a^+ a +\frac{1}{2}\big)\, \hbar\tilde{\omega} + \frac{\hbar^2\,k^2}{2\,m_r}
\end{align}
and the linear part $H_{lin}$ is given by
\begin{align}
H_{lin} = \frac{\lambda}{\sqrt{2}}\,\big(d+d^+\big) + \frac{1}{2 m^*}\, \lambda^2_o
\end{align}
where $\lambda\lambda'=\lambda^2_o/m^*$ with $\lambda_o=e E_y/\omega_c$. Equations (B8) and (B9) satisfy the
following equations:
\begin{align}
\Big[\Big(a^+ a +\frac{1}{2}\Big)\, \hbar\tilde{\omega} + \frac{\hbar^2\,k^2}{2\,m_r}\Big]\phi_n
=E_n\,\phi_n
\end{align}
and
\begin{align}
\Big[\frac{\lambda}{\sqrt{2}}\Big(d+d^+\Big) + \frac{1}{2 m^*} \lambda^2_o\Big]\phi_{\alpha}
=E_{\alpha}\,\phi_{\alpha}
\end{align}
As such, the purpose behind this problem is to determine the eigenfunctions $\phi_n$ and $\phi_{\alpha}$ and the
eigenenergies $E_n$ and $E_{\alpha}$. An analogy to the quantum harmonic oscillator now provides solutions to
Eq. (B10) to yield
\begin{align}
E_n = \Big(n +\frac{1}{2}\Big)\,\hbar\tilde{\omega} + \frac{\hbar^2 k^2}{2\, m_r},
\end{align}
the hybrid magnetoelectric subband (or Landau level, for simplicity) $n$ is the eigenvalue of $a^+ a$ and --
suppressing $\phi_k(y)$ $\big[=(1/{\scriptstyle{\sqrt{L_y}}})\,e^{i k y}$\big] throughout for the sake
of brevity --
\begin{align}
\phi_n = \frac{1}{\sqrt{n!}}\,\big(a^+\big)^n\phi_o
\end{align}
To determine the ground state $\phi_o$, we proceed as follows. Let us first  rewrite Eq. (B10) such as
\begin{align}
H\phi_n = \Big[\Big(a^+\,a +\frac{1}{2}\Big)\hbar\tilde{\omega} \Big]\phi_n = E'_n \phi_n
\end{align}
where $E'_n =E_n-\hbar^2 k^2/(2\, m_r)$. To calculate the ground state, let us first consider the average of the
Hamiltonian in Eq. (B14), i.e.,
\begin{align}
\Big(\phi_n, H\phi_n\Big)=\frac{1}{2}\hbar\tilde{\omega}\Big(\phi_n, \phi_n\Big) + \hbar\tilde{\omega}
                                          \Big(a\phi_n, a\phi_n\Big)
\end{align}
which should be real and positive since $H$ is Hermitian operator. The ground state wave function must minimize
Eq. (B15) and one can easily deduce that this holds if and only if
\begin{align}
a\,\phi_o = 0\, ,
\end{align}
which nullifies the second term on the right-hand-side of Eq. (B15) and thereby minimizes the quantity
$\Big(\phi_n, H\phi_n\Big)$. Eq. (B16) can be explicitly written as
\begin{align}
\frac{1}{2}\,\Big(z + p_z + \lambda' \Big)\,\phi_o =0
\end{align}
The differential equation (B17) has a trivial solution to be expressed, after a few simple mathematical steps, as
\begin{align}
\phi_o = N \exp\Big\{\!-\frac{x'^2}{4}-k_{\ell}\,x' +i\Big(\lambda' +\frac{y''}{2}\Big)x'\Big\},
\end{align}
where $k_{\ell}=k\,\ell_e\,(\omega_c/\tilde{\omega})$ and the coefficient $N$ to be determined through the
normalization condition $(\phi_o, \phi_o)=1$ is given by
\begin{align}
N = \frac{1}{\sqrt{\sqrt{2\pi}\,\ell_e}}\,e^{-k^2_{\ell}}
\end{align}
Thus we can write the ground state $\phi_o$ in Eq. (18) as follows:
\begin{align}
\phi_o = \frac{1}{\sqrt{\sqrt{2\pi}\,\ell_e}}
\exp\Big\{\!-\Big(\frac{x'}{2}+k_{\ell}\Big)^2 + i\Big(\lambda' +\frac{y''}{2}\Big)x'\Big\}
\end{align}
This allows us to cast the eigenfunction $\phi_n$ in Eq. (B13) in the following form:
\begin{align}
\phi_n (x, y) = \frac{1}{\sqrt{\sqrt{2\pi}\,n!\,\ell_e}}\, \big(a^+\big)^n \,
\exp\Big\{\!-\Big(\frac{x'}{2}+k_{\ell}\Big)^2 + i\Big(\lambda' +\frac{y''}{2}\Big)x'\Big\}
\end{align}
Next, let us begin with Eq. (B11), which -- with the aide of Eqs. (B3) -- assumes the form
\begin{align}
\Big[\lambda \Big(\!i\,\partial_{x'}-\frac{y''}{2}\Big)+\frac{1}{2m^*}\lambda^2_o\Big]\phi_{\alpha} =
E_{\alpha}\phi_{\alpha}\,
\end{align}
The differential equation (B22) has a trivial solution expressed, after a simple mathematical manipulation, in the
form
\begin{align}
\phi_{\alpha}=\exp\Big\{\!-i\Big(\frac{E^{'}_{\alpha}}{\lambda}+\frac{y''}{2}\Big)x' \Big\},
\end{align}
where $E^{'}_{\alpha}=E_{\alpha} - \lambda^2_o/(2\,m^*)$, and
\begin{align}
E^{'}_{\alpha}=\lambda \, \ell_e \, \alpha \Rightarrow E_{\alpha} =
              \lambda_o\, \frac{\hbar\,\alpha}{m^*} + \frac{\lambda^2_o}{2\,m^*},
\end{align}
where $\alpha \in \mathbb{R}$ labels the continuous spectrum for the $H_{lin}$. This leads us to rewrite Eq. (B23)
as
\begin{align}
\phi_{\alpha}(x, y)=\exp\Big\{\!-i\Big(\alpha\,\ell_e+\frac{y''}{2}\Big)x' \Big\}.
\end{align}
Notice that the y dependence legitimately disappears in the product $\phi_{n\alpha}(x)=\phi_n (x, y)\,\phi_{\alpha}
(x, y)$, which makes sense because this product, in fact, counts only for the dynamics along the constrained x
direction. The note in italics following Eq. (3) is worth re-emphasizing here. Equation (2) in the text is an exact
product of $\phi_{n\alpha} (x)$ and $\phi_k (y)$.




\begin{references}
\vspace{-0.55cm}
\bibitem[1]{1} H. Sakaki, Jpn. J. Appl. Phys. {\bf 19}, L735 (1980).
\bibitem[2]{2} M.S. Kushwaha, Appl. Phys. Lett. {\bf 103}, 173116 (2013).
\bibitem[3]{3} M.S. Kushwaha, Surf. Sci. Rep. {\bf 41}, 1 (2001). This is a comprehensive review article on the
               electronic, optical, and transport phenomena in the low-dimensional systems such as
               quantum-wells, -wires, -dots, and (electrically/magnetically) modulated systems.
\bibitem[4]{4} T.H. Maiman, Nature {\bf 187}(4736), 493 (1960).
\bibitem[5]{5} M.J. Connelly, {\it Semiconductor Optical Amplifiers} (Kluwer Academic, London, 2007).
\bibitem[6]{6} H. Sun, L. Yin, Z. Liu, Y. Zheng, F. Fan, S. Zhao, X. Feng, Y. Li, and C.Z. Ning, Nature
               Photonics {\bf 11}, 589 (2017).
\bibitem[7]{7} M. Dong, N.M. Mangan, J.N. Kutz, S.T. Cundiff, and H.G. Winful, IEEE J. Quantum Electron. {\bf 53},
               2500311 (2017).
\bibitem[8]{8} A.K. Mishra, O. Karni, I. Khanonkin, and G. Eisenstein, Phys. Rev. Appl. {\bf 7}, 054008 (2017).
\bibitem[9]{9} Caution: The term {\it nanowire} is an outright misnomer for a (realistic semiconducting)
               {\it quantum wire}.
\bibitem[10]{10} S.R.E. Yang and G.C. Aers, Phys. Rev. B {\bf 46}, 12456 (1992).
\bibitem[11]{11} A.R. Go\~{n}i, A. Pinczuk, J.S. Weiner, B.S. Dennis, L.N. Pfeiffer, and K.W. West, Phys. Rev.
                 Lett. {\bf 70}, 1151 (1993).
\bibitem[12]{12} Q.P. Li and S. Das Sarma, Phys. Rev. B {\bf 44}, 6277 (1991).
\bibitem[13]{13} L. Wendler and V. G. Grigoryan, Phys. Rev. B {\bf 49}, 13607 (1994).
\bibitem[14]{14} M.S. Kushwaha, Phys. Rev. B {\bf 76}, 245315 (2007).
\bibitem[15]{15} W. Kohn, Phys. Rev. {\bf 123}, 1242 (1961).
\bibitem[16]{16} Q.P. Li, K. Karrai, S.K. Yip, S. Das Sarma, and H.D. Drew, Phys. Rev. B {\bf 43}, 5151 (1991).
\bibitem[17]{17} M.S. Kushwaha, Phys. Rev. B {\bf 78}, 153306 (2008).
\bibitem[18]{18} M.S. Kushwaha, J. Appl. Phys. {\bf 109}, 106102 (2011);
\bibitem[19]{19} M.S. Kushwaha, Mod. Phys. Lett. B {\bf 28}, 1430013 (2014).
\bibitem[20]{20} M.S. Kushwaha, Euro. Phys. Lett. {\bf 123}, 34001 (2018).
\bibitem[21]{21} M.S. Kushwaha, AIP Advances {\bf 2}, 032104 (2012); {\bf 3}, 042103 (2013);
                 {\bf 6}, 035014 (2016).
\bibitem[22]{22} A. Javan, Phys. Rev. {\bf 107}, 1579 (1957).
\bibitem[23]{23} P. Y. Wen, A. F. Kockum, H. Ian, J. C. Chen, F. Nori, and I.C. Hoi, Phys. Rev. Lett. {\bf 120},
                 063603 (2018); and references therein.
\bibitem[24]{24} L.J. Wang, A. Kuzmich, and A. Dogariu, Nature {\bf 406}, 277 (2000).
\bibitem[25]{25} K.T. McDonald, Am. J. Phys. {\bf 69}, 607 (2001).
\bibitem[26]{26} G. D'Aguanno, M. Centini, M. Scalora, C. Sibilia, M. J. Bloemer, C. M. Bowden, J. W. Haus, and
                 M. Bertolotti, Phys. Rev. E {\bf 63}, 036610 (2001).
\bibitem[27]{27} A collective -- plasmon or magnetoplasmon -- excitation becomes Landau-damped and hence ceases
                 to exist as a bonafide, long-lived excitation {\it after} merging with the (respective)
                 single-particle excitation continuum in any given Q-$N$DES.
\end{references}
\end{document}